\documentclass[intlimits,twoside,a4paper]{article}
\usepackage[cp1251]{inputenc}

\usepackage[eqsecnum]{cmpj3}

\usepackage{bm}

\title[COM for 2D Hubbard and cuprates]
{The composite operator method route to
the 2D Hubbard model and the cuprates}

\author{A. Di~Ciolo\refaddr{label1}, A.~Avella\refaddr{label1,label2,label3}}
\addresses{
\addr{label1} Dipartimento di Fisica ``E.R. Caianiello'', Universit\`a degli
Studi di Salerno, I-84084 Fisciano (SA), Italy
\addr{label2} CNR-SPIN, UoS di Salerno, I-84084 Fisciano (SA), Italy
\addr{label3} Unit\`a CNISM di Salerno, Universit\`a degli Studi di Salerno, I-84084 
Fisciano (SA), Italy
}

\date{Received June 4, 2018, in final form June 28, 2018}

\begin{document}
	
\issue{2018}{21}{3}{33701}
\doinumber{10.5488/CMP.21.33701}

\maketitle

\begin{abstract}
In this review paper, we illustrate a possible route to obtain a reliable
solution of the 2D Hubbard model and an explanation for some of the
unconventional behaviours of underdoped high-$T_\text{c}$ cuprate superconductors
within the framework of the composite operator method. The latter
is described exhaustively in its fundamental philosophy, various ingredients
and robust machinery to clarify the reasons behind its successful
applications to many diverse strongly correlated systems, controversial
phenomenologies and puzzling materials.
\pacs 71.10.\textminus w, 71.10.Fd, 71.27.+a, 71.18.+y, 74.72.\textminus h,
79.60.\textminus i
\end{abstract}

\section{SCES and composite operator method}

The tale of strongly correlated electronic systems (SCES) \cite{Anderson_72,Georges_96,Hewson_97,Avella_12,Avella_13a,Avella_15a}
is deeply intertwined to that of cuprate high-$T_\text{c}$ superconductors
(HTcS) \cite{Bednorz_86,Wu_87,Timusk_99,Norman_05,Lee_06,Armitage_10,Fradkin_12,Rice_12,Avella_14a,Keimer_15,Avella_16,Novelli_17,Avella_18}
simply because the prototypical model for the former is exactly the
same as the very minimal model to describe the latter \cite{Anderson_87}:
the 2D Hubbard model \cite{Hubbard,Hubbard1964a,Hubbard1964b}. This very fact has enormously
increased the number of studies performed in the last thirty years,
that is since the discovery of HTcS, on this model and its extensions
(the Emery or $p$-$d$ model among all others \cite{Emery_87,Emery_88,Emery_94})
and derivatives (the $t$-$J$ model \cite{Oles_81,Zhang_88}, the
spin-fermion model \cite{Abanov_03}, \ldots). Consequently, also the
number of analytical and numerical methods newly developed and specially
designed to solve the 2D Hubbard model has been incredibly growing
in the last decades \cite{Avella_12,Avella_13a}. Among these methods,
several approaches employ multi-electron operators, generated by a
set of equations of motion, and the high-order Green's function (GF)
projection. In particular, in the class of \emph{operatorial} approaches
(the Hubbard approximations \cite{Hubbard,Hubbard1964a,Hubbard1964b}, an early high-order GF
approach \cite{Kuzmin_77}, the projection operator method \cite{Tserkovnikov_81,Tserkovnikov_81a},
the works of Mori \cite{Mori_65}, Rowe \cite{Rowe_68}, and Roth
\cite{Roth_69}, the spectral density approach \cite{Nolting_89},
the works of Barabanov \cite{Barabanov_01}, Val'kov \cite{Valkov_05},
and Plakida \cite{Plakida_99,Plakida_01a,Plakida_01,Plakida_06,Plakida_10},
and the cluster perturbation theory in the Hubbard-operator representation
\cite{Ovchinnikov_11}), we have been developing the composite operator
method \cite{Mancini_04,Avella_12,Avella_14,Avella_14a}.

The composite operator method (COM) has been designed and devised
and is still currently developed with the aim of providing an analytical
(operatorial) method that would seamlessly account for the natural
emergence in SCES of elementary excitations, that is quasi-particles,
whose operatorial description can only be realized in terms of fields
whose commutation relations are inherently non-canonical. Such a philosophy
requires a complete rethinking, more than a mere rewriting, of the
whole apparatus of the GF framework and of the equations of motion
(EM) formalism, which are at the basis of the operatorial approach.
As a matter of fact, the standard ingredients of the GF framework
(spectral weights, dispersions, decay rates, \ldots) need to be partly
reinterpreted in order to properly and effectively describe the system
under analysis. Moreover, new concepts --- and consequently a new
terminology --- emerge naturally when the EM formalism is applied
to such non-canonical fields.

The presence in the Hamiltonian describing the system under analysis
of strongly correlated terms~--- terms not quadratic in the \emph{canonical}
fields $\{ a_{\boldsymbol{\lambda}},b_{\boldsymbol{\mu}},\ldots\} $
describing the (quasi-)particles building up the system (electrons,
phonons, magnons, \ldots) and their internal degrees of freedom $\boldsymbol{\lambda}=\left\{ \lambda_{i}\right\} $
(spin, orbital, site, momentum, branch, \ldots) --- immediately and naturally
leads to the emergence, in the EM of the \emph{canonical} fields,
of products of the same \emph{canonical} fields which is not possible
to recast as linear combinations of the \emph{canonical} fields themselves.
Such products of \emph{canonical} fields are what we define \emph{composite}
operators (CO): they denote the very essence of strong correlations
and operatorially describe the possible excitations in the system.
Such excitations embody the strong interplay of different degrees
of freedom and/or of different families of \emph{canonical} fields
characterizing such systems. A complete basis for these new operators
(a set of operators in terms of which is possible to express all CO
emerging at the first order in the EM hierarchy as linear combinations)
contains all possible products of the \emph{number} $\{ a_{\boldsymbol{\lambda}}^{\dagger}a_{\boldsymbol{\lambda}},b_{\boldsymbol{\mu}}^{\dagger}b_{\boldsymbol{\mu}},\ldots\} $
and \emph{mixing} $\{ a_{\boldsymbol{\lambda}}^{\dagger}a_{\boldsymbol{\lambda}'},b_{\boldsymbol{\mu}}^{\dagger}b_{\boldsymbol{\mu}'},a_{\boldsymbol{\lambda}}^{\dagger}b_{\boldsymbol{\mu}},\ldots\} $
operators with the \emph{canonical} fields themselves. For instance,
the complete basis for a single site, single orbital fermionic system
is $c_{\uparrow}$, $c_{\downarrow}$, $c_{\downarrow}^{\dagger}c_{\downarrow}c_{\uparrow}$,
$c_{\uparrow}^{\dagger}c_{\uparrow}c_{\downarrow}$; the latter two
operators being the main components of the well-known Hubbard operators:
$\eta_{\sigma}=n_{\bar{\sigma}}c_{\sigma}$, $\xi_{\sigma}=\left(1-n_{\bar{\sigma}}\right)c_{\sigma}$.
In principle, at each order of the EM hierarchy, to find the complete
basis to express all possibly emerging new CO as linear combinations,
one more \emph{layer} of \emph{number} and \emph{mixing} operators
is necessary in front of each element of the previous complete basis.
The procedure to construct a complete basis at each order of the EM
hierarchy is heavily affected by the Pauli principle for purely fermionic
systems, that is by the algebraic properties of the \emph{canonical}
fields. For instance, in the previous complete basis we have no $c_{\uparrow}^{\dagger}c_{\uparrow}c_{\uparrow}$
because it is simply zero or we have that $c_{\uparrow}^{\dagger}c_{\downarrow}c_{\uparrow}=-c_{\uparrow}^{\dagger}c_{\uparrow}c_{\downarrow}$
and no \emph{mixing} operators seemed to be present while the fictitious
absence is simply related to the fact that the charge and spin degrees
of freedom on the same site are just indissolubly linked. This is
so very true that such algebraic links are at the basis of a further
fundamental ingredient of the COM {[}the Algebra Constraints (AC){]}
and unfortunately they are often overlooked leading to doomed-to-fail
approximations.

Only in purely fermionic systems with a finite number of degrees of
freedom, where the process of adding \emph{layers} is limited again
by the Pauli principle (e.g., $n_{\bar{\sigma}}n_{\bar{\sigma}}c_{\sigma}=n_{\bar{\sigma}}c_{\sigma}$),
it is possible to close self-consistently the system of EM for any
possible Hamiltonian, that is for any possible strongly correlated
term. On infinite systems, the possibility to close the hierarchy
of EM is related to (i) the effective expression of the strongly correlated
terms and (ii) if they interest those degrees of freedom which are
truly infinite (e.g., sites and momenta for a bulk system). For instance,
for the Hubbard model on the bulk, it is not possible to close the hierarchy
of the EM because the degree of freedom in which the hopping term
is diagonal, the momentum, is infinite and mixed by the Hubbard repulsion,
which is instead diagonal in the infinite site space where the hopping
term mixes. This is just the very magic of the Hubbard model!

Then, when it is not possible to close the hierarchy of the EM, one
can always choose where to stop being sure that the processes up to
those defined by the CO taken into account will be correctly described
in such an approximation. However, this is only one necessary step towards
a complete solution because to compute the GF $G$ and the correlators
$C$ of such CO under the effect of the Hamiltonian $H$ it is necessary
to calculate explicitly their \emph{weights} and \emph{overlaps} appearing
in the \emph{normalization} matrix $I$ and the \emph{connections
}among them appearing in the matrix \emph{$m$} \cite{Avella_12,Avella_14}.
This will allow one to obtain the \emph{eigenenergies} $E$ (eigenvalues
of the \emph{energy} matrix $\varepsilon$) and the \emph{spectral
weights} $\sigma$ (from the matrix $\Omega$ of eigenvectors of $\varepsilon$)
\begin{eqnarray}
&&\mathrm{i}\frac{\partial}{\partial t}\psi_{a}=\left[\psi_{a},H\right]=J_{a}=\sum_{b=1}^{n}\varepsilon_{ab}\psi_{b}+\delta J_{a}\,,\\
&&  m_{ab}=\left\langle \left\{ J_{a},\psi_{b}^{\dagger}\right\} \right\rangle, \quad I_{ab}=\left\langle \left\{ \psi_{a},\psi_{b}^{\dagger}\right\} \right\rangle, \quad\varepsilon_{ab}=\sum_{c=1}^{n}m_{ac}I_{cb}^{-1}\,,\\
&&  G_{ab}\left(\omega\right)=\mathcal{F}\left\langle \left\langle \left\{ \psi_{a},\psi_{b}^{\dagger}\right\} \right\rangle \right\rangle =\sum_{c=1}^{n}\left[\omega-\varepsilon-\Sigma\left(\omega\right)\right]_{ac}^{-1}I_{cb}\dot{\approx}\sum_{i=1}^{n}\frac{\sigma_{ab}^{\left(i\right)}}{\omega-E_{i}+\mathrm{i}\delta}\,,
\end{eqnarray}
 \begin{eqnarray}
&&  \sigma_{ab}^{(i)}=\sum_{c=1}^{n}\Omega_{ai}\Omega_{ic}^{-1}I_{cb}\,, \quad \Sigma_{ab}\left(\omega\right)=\sum_{c=1}^{n}\mathcal{F}\left\langle \left\langle \left\{ \delta J_{a},\delta J_{c}^{\dagger}\right\} \right\rangle \right\rangle _{irr}I_{cb}^{-1}\,,\\
&&  C_{ab}=\left\langle \psi_{a}\psi_{b}^{\dagger}\right\rangle =\mathcal{F}^{-1}\Big\{\int d\omega\left[1-f_{\mathrm{F}}\left(\omega\right)\right]A_{ab}\left(\omega\right)\Big\}, \quad  A_{ab}\left(\omega\right)=-\frac{1}{\piup}\Im G_{ab}\left(\omega\right),
 \end{eqnarray}
where $\psi_{a}$ belongs to the chosen set of CO $\left(\psi_{1},\ldots,\psi_{n}\right)$,
which acts as the basis in the operatorial space, $J_{a}$ is the \emph{current}
of $\psi_{a}$, $\delta J_{a}$ is the \emph{residual} current of
$\psi_{a}$ defined by $\langle \{ \delta J_{a},\psi_{b}^{\dagger}\} \rangle =0$,
$\Sigma_{ab}$ is the \emph{residual} self-energy, $A_{ab}$ is the
\emph{spectral density} and $f_{\mathrm{F}}\left(\omega\right)$ is the
Fermi function. $\mathcal{F}$ is the Fourier transform from time
to frequency domain. The latter expression on the right for $G_{ab}$
is exact if the hierarchy of EM closes on a finite number $n$ of
CO, otherwise it is the approximate expression of $G$ in the
$n$-pole approximation.

This is not the very end of the story since $I$ and $m$ usually contains
parameters to be computed. Some of these parameters are just correlators
$C_{ab}$ of the basis that can be self-consistently computed through
their expression in terms of the GF $G$ (the fluctuation-dissipation
theorem), though some others are higher-order correlators (correlators
of CO appearing at higher orders in the hierarchy of the EM) and need
to be computed in some way. COM uses the AC dictated by the algebra
obeyed by the CO in the basis, which leads to relationships between
correlators of the basis (e.g., $\langle \xi_{\sigma}\eta_{\sigma'}^{\dagger}\rangle =0$),
to fix such parameters achieving a two-fold result: (i) enforce in
the solution the AC that are not automatically satisfied contrarily
to what many people erroneously think \cite{Avella_98,Mancini_04}
and (ii) compute all unknowns in the theory.

There is still something to be computed (or neglected): the \emph{residual}
self-energy $\Sigma$. According to the physical properties under
analysis and the range of temperatures, dopings, and interactions that
we want to explore, we have to choose an approximation to compute
the \emph{residual} self-energy. In the last years, we have been using
the $n-$pole approximation \cite{Hub,Avella_98,Hub1b,Hub1c,Hub1d,Hub1e,Hub1h,Hub1i,Odashima_05,Avella_14,ttU,ttU1a,ttU1b,p-d,p-d1a,p-d1b,p-d1c,tUV,tUV1a,Plekhanov_11,Ising,Ising1a,Cuprates-NCA,Cuprates-NCA1a,Di-Ciolo_17a,Di-Ciolo_17b},
the asymptotic field approach \cite{Villani_00,Anderson,Anderson1a} and the
non-crossing approximation (NCA) \cite{Avella_07,Avella_07a,Avella_08,Avella_09,Avella_14a}.

In the last twenty years, the COM has been applied to several models
and materials: Hubbard \cite{Hub,Avella_98,Hub1b,Hub1c,Hub1d,Hub1e,Hub1h,Hub1i,Avella_03c,Krivenko_04,Odashima_05,Avella_14},
$p$-$d$ \cite{p-d,p-d1a,p-d1b,p-d1c}, $t$-$J$ \cite{Avella_02a}, $t$-$t'$-$U$
\cite{ttU,ttU1a,ttU1b}, extended Hubbard ($t$-$U$-$V$) \cite{tUV,tUV1a}, Kondo
\cite{Villani_00}, Anderson \cite{Anderson,Anderson1a}, two-orbital Hubbard
\cite{2orb,2orb1a,Plekhanov_11}, Ising \cite{Ising,Ising1a}, $J_{1}-J_{2}$ \cite{Bak_02a,J1J2,J1J21a,J1J21b,J1J21c,Avella_08a},
Hubbard-Kondo \cite{Avella_06a}, cuprates \cite{Cuprates-NCA,Cuprates-NCA1a,Avella_07,Avella_07a,Avella_08,Avella_09,Avella_14a},
etc. In this review, we will focus just on the 2D Hubbard model (section~\ref{sec:2D-Hubbard-model})
and on the best COM solutions available at the time within the $n$-pole
(section~\ref{sec:3-pole-solution}) and the NCA to the residual self-energy
(section~\ref{sec:Residual-Self-Energy-and}) approximation frameworks.
In the latter case, we will be in the condition to give a possible
explanation for some of the unconventional behaviours of underdoped
cuprate high-$T_\text{c}$ superconductors.

\section{2D Hubbard model\label{sec:2D-Hubbard-model}}

The Hamiltonian of the single-orbital 2D Hubbard model reads as 
\begin{equation}
H=-4t\sum_{\mathbf{i}}c^{\dagger}\left(i\right)\cdot c^{\alpha}\left(i\right)+U\sum_{\mathbf{i}}n_{\uparrow}\left(i\right)n_{\downarrow}\left(i\right)-\mu\sum_{\mathbf{i}}n\left(i\right),\label{eq:Ham}
\end{equation}
where $c^{\dagger}\left(i\right)=(c_{\uparrow}^{\dagger}\left(i\right),c_{\downarrow}^{\dagger}\left(i\right))$
is the electronic field operator in spinorial notation and Heisenberg
picture ($i=\left(\mathbf{i},t_{i}\right)$). $\cdot$ and $\otimes$
stand for the inner (scalar) and the outer products, respectively,
in spin space. $\mathbf{i}$ is a vector of the two-dimensional square
Bravais lattice, $n_{\sigma}\left(i\right)=c_{\sigma}^{\dagger}\left(i\right)c_{\sigma}\left(i\right)$
is the particle density operator for spin $\sigma$ at site $\mathbf{i}$,
$n\left(i\right)=\sum_{\sigma}n_{\sigma}\left(i\right)=c^{\dagger}\left(i\right)\cdot c\left(i\right)$
is the total particle density operator at site $\mathbf{i}$, $\mu$
is the chemical potential, $t$ is the hopping integral and the energy
unit hereafter, $U$ is the Coulomb on-site repulsion and $\alpha_{\mathbf{ij}}$
is the projector on the nearest-neighbor sites with Fourier transform
$\alpha(\mathbf{k})=\mathcal{F}_{\mathbf{k}}\alpha_{\mathbf{ij}}=\frac{1}{2}[\cos(k_{x}a)+\cos(k_{y}a)]$.
For any operator $\psi\left(i\right)$, we use the notation $\psi^{\kappa}\left(i\right)=\sum_{\mathbf{j}}\kappa_{\mathbf{ij}}\psi\left(\mathbf{j},t_{i}\right)$
where $\kappa_{\mathbf{ij}}$ can be any function of the two sites
$\mathbf{i}$ and $\mathbf{j}$.

\newpage

\section{3-pole solution\label{sec:3-pole-solution}}

\subsection{Basis and equations of motion\label{sec:Basis}}

Following the COM prescription \cite{Mancini_04,Avella_11a,Avella_14,Avella_14a},
we have chosen a basic field and, in particular, we have selected
the following composite triplet field operator $\psi\left(i\right)=(\xi^{\dagger}\left(i\right),\eta^{\dagger}\left(i\right),c_{s}^{\dagger}\left(i\right))$
where $\eta\left(i\right)=n\left(i\right)c\left(i\right)$ and $\xi\left(i\right)=c\left(i\right)-\eta\left(i\right)$,
the Hubbard operators, satisfy the following EM: 
\begin{equation}
\mathrm{i}\frac{\partial}{\partial t}\xi\left(i\right)=-\mu\xi\left(i\right)-4tc^{\alpha}\left(i\right)-4t\pi\left(i\right),\qquad\mathrm{i}\frac{\partial}{\partial t}\eta\left(i\right)=\left(U-\mu\right)\eta\left(i\right)+4t\pi\left(i\right)\label{eq:em-xi},
\end{equation}
where the higher-order composite field $\pi\left(i\right)=\frac{1}{2}n_{\mu}(i)\sigma^{\mu}\cdot c^{\alpha}\left(i\right)+c^{\dagger\alpha}\left(i\right)\cdot c\left(i\right)\otimes c\left(i\right)$
and $n_{\mu}\left(i\right)=c^{\dagger}\left(i\right)\cdot\sigma_{\mu}\cdot c\left(i\right)$
is the charge- ($\mu=0$) and spin- ($\mu=1,2,3=k$) density operator,
$\sigma_{\mu}=\left(1,\vec{\sigma}\right)$, $\sigma^{\mu}=\left(-1,\vec{\sigma}\right)$,
$\sigma_{k}$ with $\left(k=1,2,3\right)$ are the Pauli matrices.

The third operator in the basis, $c_{s}\left(i\right)$, is chosen
proportional to the \emph{spin} component of $\pi\left(i\right)$:
$c_{s}\left(i\right)=n_{k}\left(i\right)\sigma_{k}\cdot c^{\alpha}\left(i\right)$.
Accordingly, we define $\bar{\pi}\left(i\right)=\pi\left(i\right)-\frac{1}{2}c_{s}\left(i\right)$.
The use of $c_{s}\left(i\right)$ will highlight the most relevant
physics as we do expect spin fluctuations to be the most relevant
ones. For further details, such as the EM satisfied by the field $c_{s}\left(i\right)$,
see reference~\cite{Avella_14}.

\subsection{Normalization $I$ matrix\label{sec:I}}

In the paramagnetic and homogeneous case, the entries of the normalization
matrix $I(\mathbf{k})=$ \linebreak $\mathcal{F}_{\mathbf{k}}\langle \{ \psi\left(\mathbf{i},t\right),\psi^{\dagger}\left(\mathbf{j},t\right)\} \rangle $
have the following expressions
\begin{align}
I_{11}\left(\mathbf{k}\right) & =I_{11}=1-\frac{n}{2}\,,\qquad I_{12}\left(\mathbf{k}\right)=0\,,\qquad I_{22}\left(\mathbf{k}\right)=I_{22}=\frac{n}{2}\label{eq:I11k}\,, \\
I_{13}\left(\mathbf{k}\right) & =3C_{\xi c}^{\alpha}+\frac{3}{2}\alpha\left(\mathbf{k}\right)\chi_{s}^{\alpha}\,,\qquad I_{23}\left(\mathbf{k}\right)=3C_{\eta c}^{\alpha}-\frac{3}{2}\alpha\left(\mathbf{k}\right)\chi_{s}^{\alpha}\label{eq:I13k}\,, \\
I_{33}\left(\mathbf{k}\right) & =4C_{c_{s}c}^{\alpha}+\frac{3}{2}C_{\eta\eta}+3\alpha\left(\mathbf{k}\right)\left(f_{s}+\frac{1}{4}C_{cc}^{\alpha}\right)+\frac{3}{2}\beta\left(\mathbf{k}\right)\chi_{s}^{\beta}+\frac{3}{4}\eta\left(\mathbf{k}\right)\chi_{s}^{\eta}\,,
\end{align}
where $n=\left\langle n\left(i\right)\right\rangle $ is the filling,
$\chi_{s}^{\kappa}=\frac{1}{3}\langle n_{k}^{\kappa}\left(i\right)n_{k}\left(i\right)\rangle $
is the spin-spin correlation function at distances determined by the
projector $\kappa$ and $f_{s}=\frac{1}{3}\langle c^{\dagger}\left(i\right)\cdot\sigma_{k}\cdot c^{\alpha}\left(i\right)n_{k}^{\alpha}\left(i\right)\rangle $
is a higher-order (up to three different sites are involved) spin-spin
correlation function. We have also introduced the following definitions,
which are based on those related to the correlation functions of the
fields of the basis: $C_{\phi\varphi}=\langle \phi_{\sigma}\left(i\right)\varphi_{\sigma}^{\dagger}\left(i\right)\rangle $
and $C_{\phi\varphi}^{\kappa}=\langle \phi_{\sigma}^{\kappa}\left(i\right)\varphi_{\sigma}^{\dagger}\left(i\right)\rangle $,
where no summation over $\sigma$ is intended. $\beta\left(\mathbf{k}\right)$
and $\eta\left(\mathbf{k}\right)$ are the projectors onto the second-nearest-neighbor
sites along the main diagonals and the main axes of the lattice, respectively.

\subsection{$m$-matrix\label{sec:m}}

In the same conditions, the entries of the matrix $m\left(\mathbf{k}\right)=\mathcal{F}_{\mathbf{k}}\left\langle \left\{ \mathrm{i}\frac{\partial}{\partial t}\psi\left(\mathbf{i},t\right),\psi^{\dagger}\left(\mathbf{j},t\right)\right\} \right\rangle $
have the following expressions
\begin{align}
m_{11}\left(\mathbf{k}\right) & =-\mu I_{11}-4t\left[\Delta+\left(p+I_{11}-I_{22}\right)\alpha\left(\mathbf{k}\right)\right]\label{eq:m11k}, \\ 
\nonumber\\
m_{12}\left(\mathbf{k}\right) & =4t\left[\Delta+\left(p-I_{22}\right)\alpha\left(\mathbf{k}\right)\right]\label{eq:m12k}, \\
\nonumber\\
m_{13}\left(\mathbf{k}\right) & =-\left[\mu+4t\alpha\left(\mathbf{k}\right)\right]I_{13}\left(\mathbf{k}\right)-4t\alpha\left(\mathbf{k}\right)I_{23}\left(\mathbf{k}\right)-2tI_{33}\left(\mathbf{k}\right)-4tI_{\bar{\pi}c_{s}}\left(\mathbf{k}\right), \\
\nonumber\\
m_{22}\left(\mathbf{k}\right) & =\left(U-\mu\right)I_{22}-4t\left[\Delta+p\alpha\left(\mathbf{k}\right)\right]\label{eq:m22k}, \\
\nonumber\\
m_{23}\left(\mathbf{k}\right) & =\left(U-\mu\right)I_{23}\left(\mathbf{k}\right)+2tI_{33}\left(\mathbf{k}\right)+4t\alpha\left(\mathbf{k}\right)I_{\bar{\pi}c_{s}}^{\alpha}\label{eq:m23k}, 
\end{align}
\begin{align}
m_{33}\left(\mathbf{k}\right) & \cong-\mu I_{33}\left(\mathbf{k}\right)+\bar{m}_{33}^{0}+\alpha\left(\mathbf{k}\right)\bar{m}_{33}^{\alpha}\,,\label{eq:m33k}
\end{align}
where $\Delta=C_{\xi\xi}^{\alpha}-C_{\eta\eta}^{\alpha}$ is the difference
between upper and lower intra-Hubbard-subband contributions to the
kinetic energy and $p=\frac{1}{4}(\chi_{0}^{\alpha}+3\chi_{s}^{\alpha})-\chi_{p}^{\alpha}$
is a combination of the nearest-neighbor charge-charge $\chi_{0}^{\alpha}=\left\langle n^{\alpha}\left(i\right)n\left(i\right)\right\rangle $,
spin-spin $\chi_{s}^{\alpha}$ and pair-pair $\chi_{p}^{\alpha}=\langle [c_{\uparrow}\left(i\right)c_{\downarrow}\left(i\right)]^{\alpha}c_{\downarrow}^{\dagger}\left(i\right)c_{\uparrow}^{\dagger}\left(i\right)\rangle $
correlation functions. We can avoid cumbersome and somewhat meaningless
calculations (they will lead to the appearance of many unknown higher-order
correlation functions) by restricting $m_{33}\left(\mathbf{k}\right)$
just to the local and the nearest-neighbor terms: $\bar{m}_{33}^{0}$
and $\bar{m}_{33}^{\alpha}$. Given the overall choice of cutting
cubic harmonics higher than the nearest-neighbor ones, for the sake
of consistency, we also neglected the $\beta\left(\mathbf{k}\right)$
and $\eta\left(\mathbf{k}\right)$ terms in $I_{33}\left(\mathbf{k}\right)$.
We checked that this latter simplification does not lead to any appreciable
difference: within the explored paramagnetic solution, $\chi_{s}^{\beta}$
and $\chi_{s}^{\eta}$ have not very significative values.

By checking systematically all operatorial relations existing among
the fields of the basis, we can recognize the following AC
\begin{align}
 & C_{\xi\xi}=1-n+D\,,\qquad C_{\xi\eta}=0\,,\qquad C_{\eta\eta}=\frac{n}{2}-D\,,\label{eq:Cxixi} \\
 & C_{\xi c_{s}}=3C_{\xi c}^{\alpha}\,,\qquad C_{\eta c_{s}}=0\,,\label{eq:Cxics}
\end{align}
where $D=\langle n_{\uparrow}\left(i\right)n_{\downarrow}\left(i\right)\rangle $
is the double occupancy. These relations lead to the following very
relevant ones: $n=2(1-C_{\xi\xi}-C_{\eta\eta})$ and $D=1-C_{\xi\xi}-2C_{\eta\eta}$.

On the other hand, we can compute $\chi_{0}^{\alpha}$, $\chi_{s}^{\alpha}$,
$\chi_{p}^{\alpha}$ and $f_{s}$ by operatorial projection, which
is equivalent to the well-established one-loop approximation \cite{Theory,Theory1a,Avella_11a}
for same-time correlations functions 
\begin{align}
\chi_{0}^{\alpha} & \approx n^{2}-2\frac{I_{11}\big(C_{c\eta}^{\alpha}\big)^{2}+I_{22}\big(C_{c\xi}^{\alpha}\big)^{2}}{C_{\eta\eta}}\,,\quad\chi_{s}^{\alpha}\approx-2\frac{I_{11}\big(C_{c\eta}^{\alpha}\big)^{2}+I_{22}\big(C_{c\xi}^{\alpha}\big)^{2}}{2I_{11}I_{22}-C_{\eta\eta}}\,,\quad\chi_{p}^{\alpha}\approx\frac{C_{c\xi}^{\alpha}C_{\eta c}^{\alpha}}{C_{\eta\eta}}\,,\label{eq:chia0}\\
f_{s} & \approx-\frac{1}{2}C_{c\xi}^{\alpha}-\frac{3}{4}\chi_{s}^{\alpha}\left(\frac{C_{c\xi}^{\alpha}}{I_{11}}-\frac{C_{c\eta}^{\alpha}}{I_{22}}\right)-2\frac{C_{c\xi}^{\alpha}}{I_{11}}\left(C_{c\xi}^{\alpha^{2}}-\frac{1}{4}C_{c\xi}\right)-2\frac{C_{c\eta}^{\alpha}}{I_{22}}\left(C_{c\eta}^{\alpha^{2}}-\frac{1}{4}C_{c\eta}\right).\nonumber 
\end{align}
As a matter of fact, the energy matrix $\varepsilon(\mathbf{k})=m(\mathbf{k})I^{-1}(\mathbf{k})$
is sure to have real eigenvalues if the normalization matrix $I\left(\mathbf{k}\right)$
is semi-positive.\footnote{The product of two symmetric matrices has real eigenvalues if one of the two is
  semi-positive, that is, it has positive or null eigenvalues.} Then, the presence of $\chi_{s}^{\alpha}$
and $f_{s}$ in the normalization matrix $I\left(\mathbf{k}\right)$
 imposes a special care in evaluating their values. Accordingly, we
have decided to avoid using AC to fix them, and to fix $\chi_{0}^{\alpha}$
and $\chi_{p}^{\alpha}$ for the sake of consistency. AC, in the attempt
to preserve the operatorial relations they stem from, can lead to
values of the unknowns slightly off their physical bounds in the spirit
of using them as mere parameters to achieve the ultimate task of satisfying
the operatorial algebra at the level of averages.

\begin{figure}[!h]
	\begin{centering}
		\begin{tabular}{ccc}
			\includegraphics[width=0.25\textwidth]{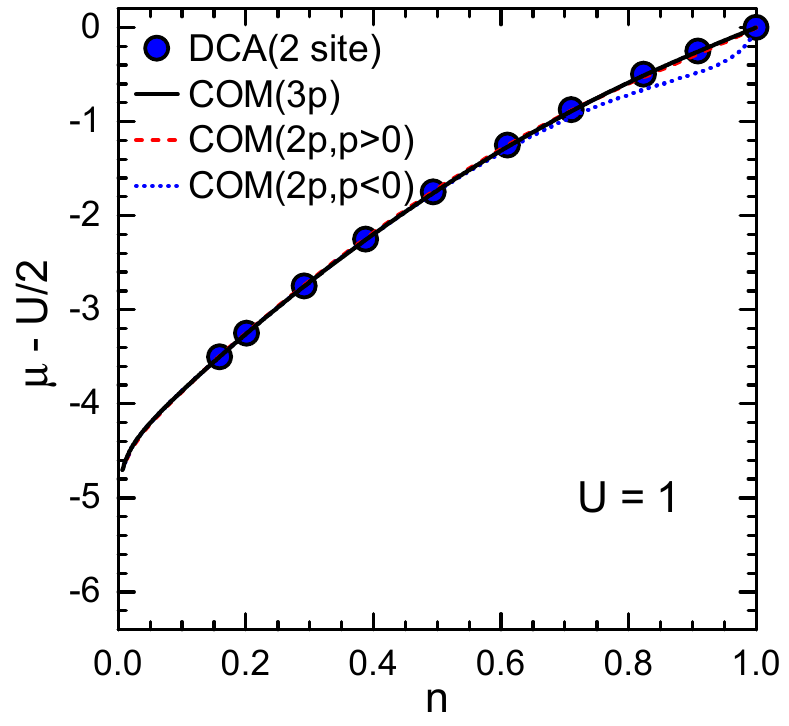} & \includegraphics[width=0.25\textwidth]{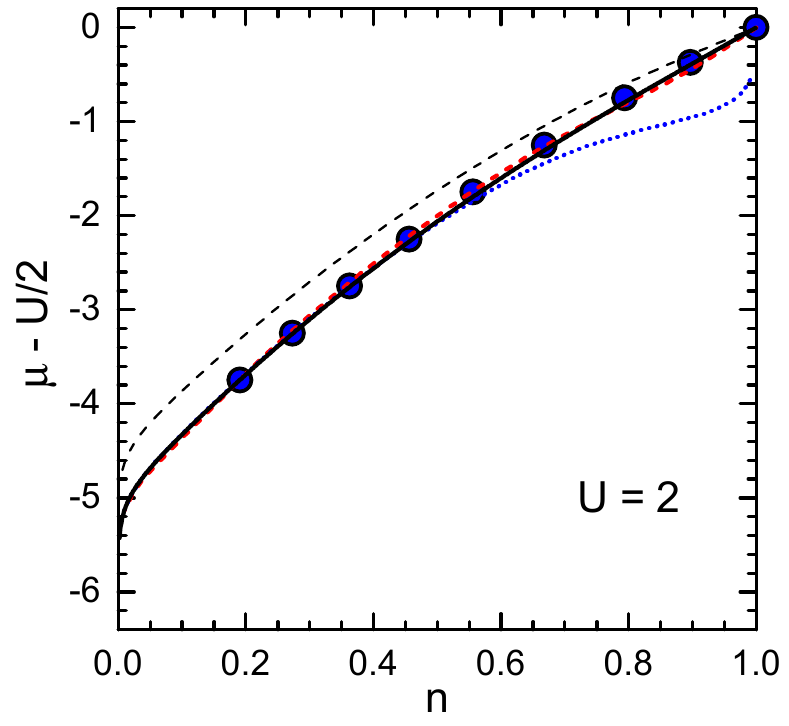} & \includegraphics[width=0.25\textwidth]{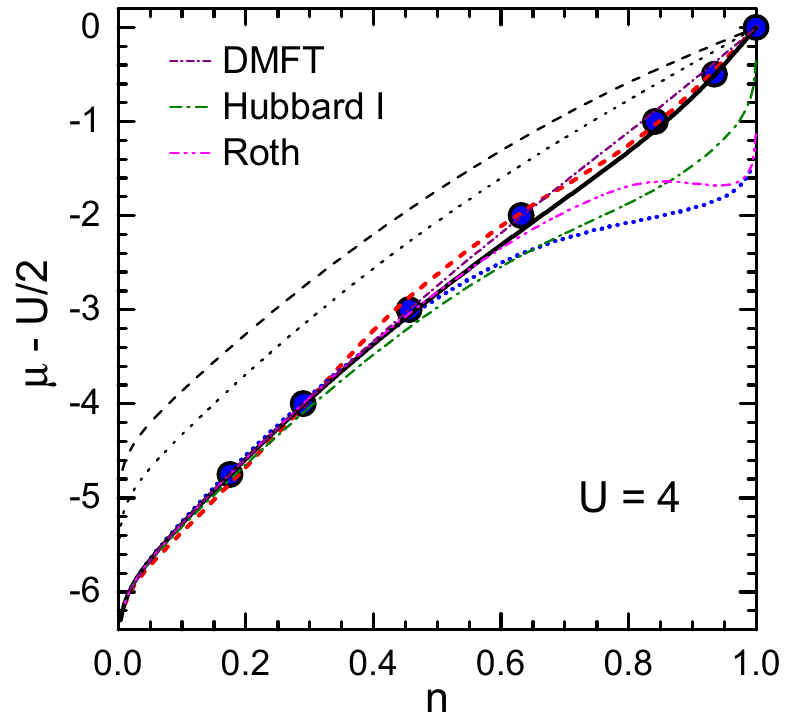}\tabularnewline
			\includegraphics[width=0.25\textwidth]{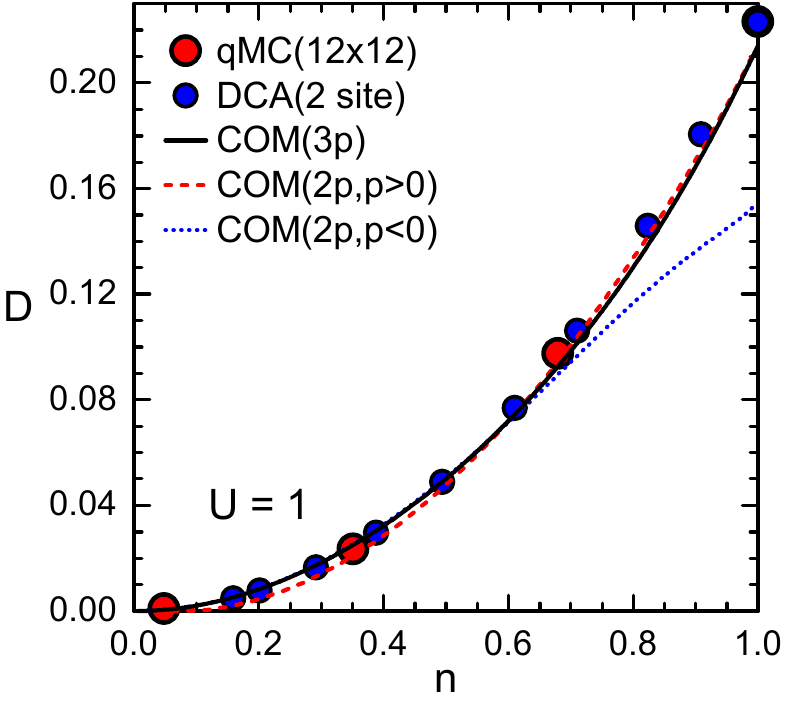} & \includegraphics[width=0.25\textwidth]{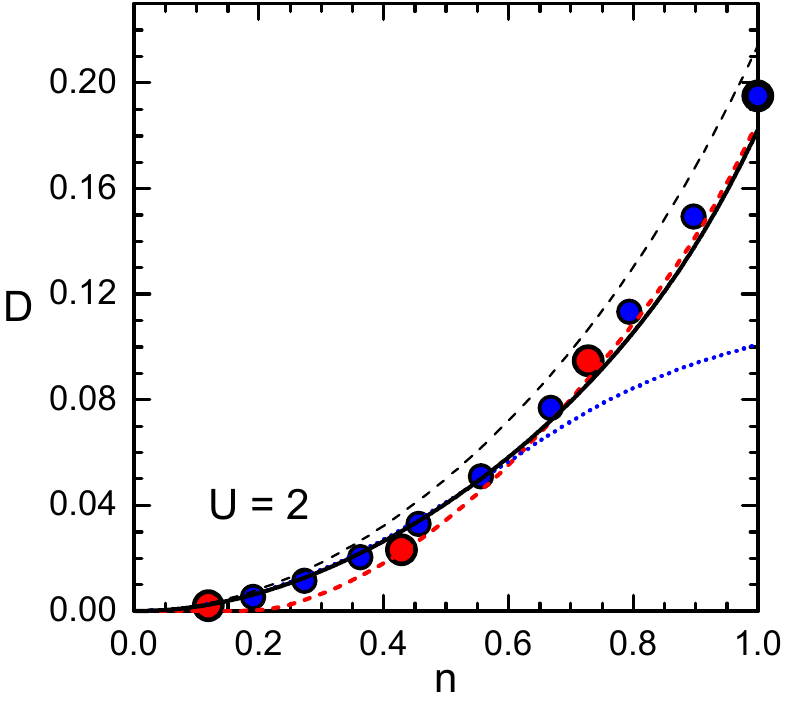} & \includegraphics[width=0.25\textwidth]{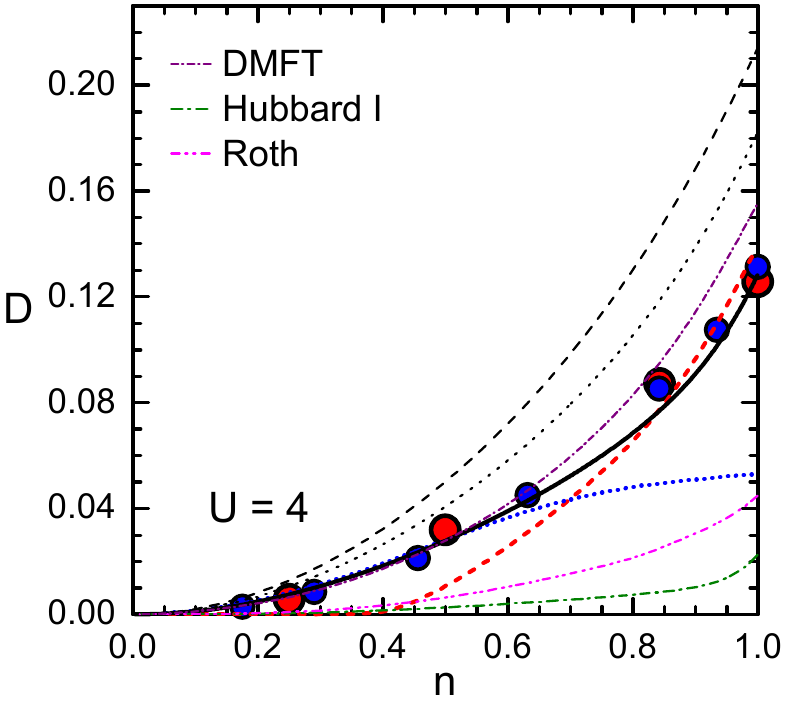}\tabularnewline
		\end{tabular}
		\par\end{centering}
	\caption{(Colour online) Scaled chemical potential $\mu-{U}/{2}$ (top row) and double
		occupancy $D$ (bottom row) as functions of the filling $n$ for $U=1$
		(left-hand column), $2$ (central column) and $4$ (right-hand column) and $T={1}/{6}$
		for COM(3p) (black lines), COM(2p,$p>0$) (dashed red line) and COM(2p,$p<0$)
		(dotted blue line). COM results are compared with $12\times12$-site
		qMC \cite{Moreo_90} and $2$-site DCA \cite{Sangiovanni} numerical
		data (red and blue circles, respectively) as well as with the results
		of Hubbard I (dot-dashed green line) and Roth (dot-dot-dashed magenta
		line) methods (only at $U=4$). The thin black dashed and dotted lines
		in central and right-hand columns are COM(3p) results for $U=1$ and $U=2$,
		respectively. From \cite{Avella_14} with kind permission of The European
		Physical Journal (EPJ).\label{fig:mu_D}}
\end{figure}

In figure~\ref{fig:mu_D}, we report the behaviour of the chemical
potential $\mu$ and of the double occupancy~$D$ as functions of
the filling $n$ for $U=1$, $2$ and $4$ and $T={1}/{6}$.
The COM $3$-pole solution [COM(3p)] is clearly in very good agreement
for all values of $U$ reported in the whole range of filling $n$
with the $12\times12$-site qMC \cite{Moreo_90} and $2$-site DCA
\cite{Sangiovanni} numerical data. COM(3p) also catches the change
of concavity in the chemical potential in proximity of half filling
between $U=1,2$ and $U=4$ [figure~\ref{fig:mu_D} (top row)] reported
by the DCA data. Also the COM $2$-pole solution with the $p$ parameter
positive {[}COM(2p, $p>0$){]} manages, but not COM(2p, $p<0$), Hubbard
I and Roth solutions, which always present the same concavity. $U=4$
already drives quite strong electronic correlations, and the chemical
potential shows a tendency towards a gap opening at $n=1$. The double
occupancy $D$ [figure~\ref{fig:mu_D} (bottom row)] reports a clear
change of correlation-strength regime between $U=1,2$ and $U=4$:
it goes from a parabolic-like behaviour resembling the non-interacting
one (${n^{2}}/{4}$) at $U=1,2$ [figure~\ref{fig:mu_D} (bottom-left/central
panels)] to a behaviour presenting a change of slope on approaching
half filling at $U=4$ [figure~\ref{fig:mu_D} (bottom-right panel)].
COM(3p) only describes correctly these features. COM(3p) evidently
has [see figure~\ref{fig:mu_D} (bottom-central/right panels)] the
capability to correctly interpolate between the two COM(2p) solutions
sticking to COM(2p, $p<0$) at low-intermediate values of filling
and even improving on COM(2p, $p>0$) at larger values of filling.

\section{Residual self-energy and cuprate superconductors\label{sec:Residual-Self-Energy-and}}

\subsection{Theory\label{sec2.1}}

In this case, we choose as basic field just the composite doublet
field operator $\psi\left(i\right)=\left(\xi^{\dagger}\left(i\right),\eta^{\dagger}\left(i\right)\right)$.
By considering the two-time thermodynamic GF \cite{Bogoliubov_59,Zubarev_60,Zubarev_74},
we define the retarded GF $G(i,j)=\langle R[\psi(i)\psi^{\dagger}(j)]\rangle$ $=\theta(t_{i}-t_{j})\langle\{\psi(i),\,\psi^{\dagger}(j)\}\rangle$
that satisfies the following equation 
\begin{equation}
\Lambda(\partial_{i})G(i,j)\Lambda^{\dagger}(\overleftarrow{\partial}_{j})=\Lambda(\partial_{i})G_{0}(i,j)\Lambda^{\dagger}(\overleftarrow{\partial}_{j})+\langle R[\delta J(i)\delta J^{\dagger}(j)]\rangle\,,\label{3.14}
\end{equation}
where the derivative operator $\Lambda(\partial_{i})=\mathrm{i}\frac{\partial}{\partial t_{i}}-\varepsilon(-\mathrm{i}\nabla_{i})$
and the propagator $G^{0}(i,j)$ is defined by the equation $\Lambda(\partial_{i})G^{0}(i,j)=\mathrm{i}\delta(t_{i}-t_{j})I(i,j)$.
By introducing the Fourier transform, equation (\ref{3.14}) can be
formally solved as 
\begin{equation}
G(\mathbf{k},\omega)=\frac{1}{\omega-\varepsilon(\mathbf{k})-\Sigma(\mathbf{k},\omega)}I(\mathbf{k})\,,\label{3.19}
\end{equation}
where the self-energy $\Sigma(\mathbf{k},\omega)$ has the expression
$\Sigma(\mathbf{k},\omega)=B_{irr}(\mathbf{k},\omega)I^{-1}(\mathbf{k})$
with $B(\mathbf{k},\omega)=$ \linebreak $\mathcal{F}\langle R[\delta J(i)\delta J^{\dagger}(j)]\rangle$.
Equation (\ref{3.19}) is the Dyson equation for composite fields
and represents the starting point for a perturbative calculation in
terms of the propagator $G^{0}(\mathbf{k},\omega)=\frac{1}{\omega-\varepsilon(\mathbf{k})}I(\mathbf{k})$,
which corresponds to the $2$-pole approximation and can be easily
obtained by the calculations in the previous sections.

The calculation of the self-energy $\Sigma(\mathbf{k},\omega)$ requires
the calculation of the higher-order propagator $B(\mathbf{k},\omega)$.
We shall compute this quantity by using the non-crossing approximation
(NCA). By neglecting the pair term $c(i)c^{\dagger\alpha}(i)c(i)$,
the source $J(i)$ can be written as $J(\mathbf{i},t)=\sum_{\mathbf{j}}a(\mathbf{i,j},t)\psi(\mathbf{j},t)$
where 
\begin{eqnarray}
&&a_{11}(\mathbf{i,j},t)=-\mu\delta_{\mathbf{ij}}-4t\alpha_{\mathbf{ij}}-2t\sigma^{\mu}n_{\mu}(i)\alpha_{\mathbf{ij}}\,,\quad a_{12}(\mathbf{i,j},t)=-4t\alpha_{\mathbf{ij}}-2t\sigma^{\mu}n_{\mu}(i)\alpha_{\mathbf{ij}}\,, \nonumber \\ 
&&a_{21}(\mathbf{i,j},t)=2t\sigma^{\mu}n_{\mu}(i)\alpha_{\mathbf{ij}}\,,\quad a_{22}(\mathbf{i,j},t)=(U-\mu)\delta_{\mathbf{ij}}+2t\sigma^{\mu}n_{\mu}(i)\alpha_{\mathbf{ij}}.
\label{3.31}
\end{eqnarray}
Then, for the calculation of $B_{irr}(i,j)=\langle R[\delta J(i)\delta J^{\dagger}(j)]\rangle_{irr}$,
we approximate $\delta J(\mathbf{i},t)\approx\sum_{\mathbf{j}}[a(\mathbf{i,j},t)-$ \linebreak $\langle a(\mathbf{i,j},t)\rangle]\psi(\mathbf{j},t)$
and, therefore, $B_{irr}(i,j)=4t^{2}F(i,j)(1-\sigma_{1})$ where we
defined $F(i,j)=$ \linebreak$\langle R[\sigma^{\mu}\delta n_{\mu}(i)c^{\alpha}(i)c^{\dagger\alpha}(j)\delta n_{\lambda}(j)\sigma^{\lambda}]\rangle$
with $\delta n_{\mu}(i)=n_{\mu}(i)-\langle n_{\mu}(i)\rangle$. The
self-energy is then written as 
\begin{equation}
\Sigma(\mathbf{k},\omega)=4t^{2}F(\mathbf{k},\omega)\left(\begin{array}{cc}
I_{11}^{-2} & -I_{11}^{-1}I_{22}^{-1} \vspace{1mm}\\
-I_{11}^{-1}I_{22}^{-1} & I_{22}^{-2}
\end{array}\right).\label{3.35}
\end{equation}

In order to calculate the retarded function $F(i,j)$, first we use
the spectral theorem to express 
\begin{equation}
F(i,j)=\frac{\ri}{2\piup}\int_{-\infty}^{+\infty} \rd\omega\mathrm{e}^{-\mathrm{i}\omega(t_{i}-t_{j})}\frac{1}{2\piup}\int_{-\infty}^{+\infty}\rd\omega'\frac{1+\mathrm{e}^{-\beta\omega'}}{\omega-\omega'+\mathrm{i}\varepsilon}C(\mathbf{i-j},\omega')\,,\label{3.36}
\end{equation}
where $C(\mathbf{i-j},\omega')$ is the correlation function $C(i,j)=\langle\sigma^{\mu}\delta n_{\mu}(i)c^{\alpha}(i)c^{\dagger\alpha}(j)\delta n_{\lambda}(j)\sigma^{\lambda}\rangle$.
Next, we use the NCA and approximate $\langle\sigma^{\mu}\delta n_{\mu}(i)c^{\alpha}(i)c^{\dagger\alpha}(j)\delta n_{\lambda}(j)\sigma^{\lambda}\rangle\approx\langle\delta n_{\mu}(i)\delta n_{\mu}(j)\rangle\langle c^{\alpha}(i)c^{\dagger\alpha}(j)\rangle$.
By means of this decoupling and using again the spectral theorem, we
finally have 
\begin{eqnarray}
F(\mathbf{k},\omega)&=&\frac{1}{\piup}\int_{-\infty}^{+\infty}\rd\omega'\frac{1}{\omega-\omega'+\mathrm{i}\delta}\frac{a^{2}}{(2\piup)^{3}}\int \rd^{2}p\rd\Omega\alpha^{2}(p) \nonumber\\
&\times&\left[\tanh\frac{\beta\Omega}{2}+\coth\frac{\beta(\omega'-\Omega)}{2}\right]\Im[G_{cc}(\mathbf{p},\Omega)]\Im[\chi(\mathbf{k-p},\omega'-\Omega)]\,,
\label{3.39}
\end{eqnarray}
where $G_{cc}(\mathbf{k},\omega)=\sum_{a,b=1}^{2}G_{ab}(\mathbf{k},\omega)$
is the retarded electronic GF and $\chi(\mathbf{k},\omega)=\sum_{\mu}\mathcal{F}\langle R[\delta n_{\mu}(i)\delta n_{\mu}(j)]\rangle $
is the total charge and spin dynamical susceptibility. The result
(\ref{3.39}) shows that the self-energy calculation requires the
knowledge of the bosonic propagator. Remarkably, up to this point,
the system of equations for the GF and the \emph{anomalous }self-energy
is similar to the one derived in the two-particle self-consistent
approach (TPSC) \cite{Vilk_95,Tremblay_06}, the DMFT$+\Sigma$ approach
\cite{Sadovskii_01,Sadovskii_05,Kuchinskii_05,Kuchinskii_06,Kuchinskii_06a}
and a Mori-like approach by Plakida and coworkers \cite{Plakida_01,Plakida_06,Plakida_10}.
It would be the way to compute the dynamical spin and charge susceptibilities
to be completely different because, instead of relying on a phenomenological
model and neglecting the charge susceptibility like these approaches
do, we will use a self-consistent two-pole approximation \cite{Hub1d}.

Finally, the electronic GF $G(\mathbf{k},\omega)$ is computed through
the following self-consistency scheme: we first compute $G^{0}(\mathbf{k},\omega)$
and $\chi_{\mu}(\mathbf{k},\omega)$ in the two-pole approximation,
then $\Sigma(\mathbf{k},\omega)$ and consequently $G(\mathbf{k},\omega)$.
Finally, we check how much the fermionic parameters ($\mu$, $\Delta$,
and $p$) changed and decide whether to stop or to continue by computing
new $\chi_{_{\mu}}(\mathbf{k},\omega)$ and $\Sigma(\mathbf{k},\omega)$
after $G(\mathbf{k},\omega)$.

\subsection{Results\label{sec:Results}}

We intend to characterize some electronic properties by computing
the spectral function $A(\mathbf{k},\omega)=-\frac{1}{\piup}\Im[G_{cc}(\mathbf{k},\omega)]$
and the density of states per spin $N(\omega)=\frac{1}{(2\piup)^{2}}\int \rd^{2}kA(\mathbf{k},\omega)$.
We also investigate the electronic self-energy $\Sigma_{cc}(\mathbf{k},\omega)$,
defined through the equation 
\begin{equation}
G_{cc}(\mathbf{k},\omega)=\frac{1}{\omega-\epsilon_{0}(\mathbf{k})-\Sigma_{cc}(\mathbf{k},\omega)}\,,
\end{equation}
where $\epsilon_{0}(\mathbf{k})=-\mu-4t\alpha(\mathbf{k})$ is the
non-interacting dispersion, and introduce the quantity $r(\mathbf{k})=\epsilon_{0}(\mathbf{k})+\Sigma'_{cc}(\mathbf{k},\omega=0)$
that determines the Fermi surface locus in the momentum space, $r(\mathbf{k})=0$,
in a Fermi liquid. The actual Fermi surface (or its relic in a non-Fermi-liquid)
is given by the relative maxima of $A(\mathbf{k},\omega=0)$, which
takes into account, on equal footing, both $\Sigma'_{cc}(\mathbf{k},\omega)$
and $\Sigma''_{cc}(\mathbf{k},\omega)$, and is directly related,
within the sudden approximation and forgetting any selection rules,
to the effective ARPES measurements.

\begin{figure}[!t]
\noindent \centering{}\includegraphics[height=0.255\textheight]{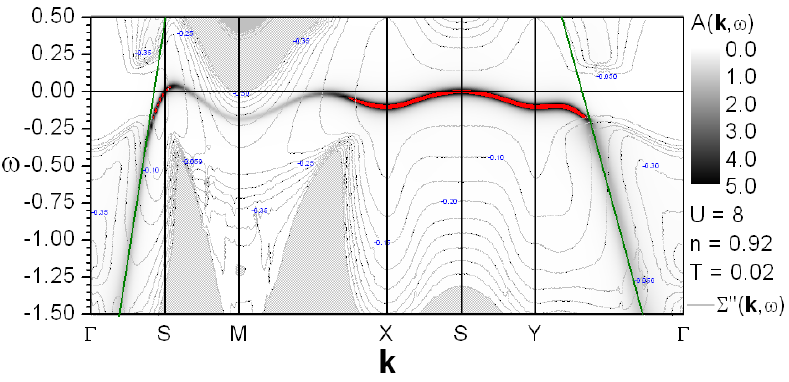}
\caption{(Colour online) Spectral function $A(\mathbf{k},\omega)$ close to
the chemical potential ($\omega=0$) along the principal directions
[$\Gamma=(0,\,0)\to S=(\piup/2,\,\piup/2)\to M=(\piup,\,\piup)$, $M\to X=(\piup,\,0)$,
$X\to Y=(0,\,\piup)$ and $Y\to\Gamma$] for $U=8$, ${\normalcolor T=0.02}$
and $n=0.92$. From \cite{Avella_14a}.\label{fig10}}
\end{figure}

\subsubsection{Spectral function and dispersion\label{sec:Dispersion}}

The electronic dispersion, or better its relic in a strongly correlated
system, can be obtained through the relative maxima of $A(\mathbf{k},\omega)$.
In figure~\ref{fig10}, we show the latter, in scale of grays (red
is for the above-scale values), along the principal directions [$\Gamma=(0,\,0)\to S=(\piup/2,\,\piup/2)\to M=(\piup,\,\piup)$,
$M\to X=(\piup,\,0)$, $X\to Y=(0,\,\piup)$ and $Y\to\Gamma$] for $U=8$,
$T=0.02$ and $n=0.92$. The light gray lines and uniform areas are
labeled with the values of $\Sigma''(\mathbf{k},\omega)$. The dark
green lines are guides to the eye, signaling the direction of the
dispersion just \emph{before} the visible kink separating the black
and the red areas of the dispersion. The latter is well-defined (red
areas) only in the regions where $\Sigma''(\mathbf{k},\omega)$ is
zero or almost negligible. In the regions where $\Sigma''(\mathbf{k},\omega)$
is instead finite, $A(\mathbf{k},\omega)$ assumes very low values,
very difficult to be detected by ARPES, which, accordingly, would
report only the red areas in the picture. This is crucial to understand
the experimental evidences on the Fermi surface in the underdoped
regime and to explain ARPES findings together with those of quantum
oscillations measurements.

The dispersion loses significance close to $M$ where $A(\mathbf{k},\omega)$
loses weight as $\Sigma''(\mathbf{k},\omega)$ increases: correspondingly,
$\chi_{3}\left(\mathbf{k},\omega\right)$ is strongly peaked at $M$
due to the strong antiferromagnetic correlations in the system. The
bandwidth reduces from $4t$ to values of the order $J={4t^{2}}/{U}=0.5t$,
as expected for the dispersion of few holes in a strong antiferromagnetic
background. The characteristics of the dispersion are also compatible
with this scenario, such as the sequence of minima and maxima, actually
driven by the doubling of the Brillouin zone induced by the strong
antiferromagnetic correlations, as well as the dynamical formation
of a $t'$ diagonal hopping indicated by the pronounced warping of
the dispersion along the $X\to Y$ direction. Moreover, remarkably,
the dispersion at $X$($Y$) coming from both $\Gamma$ and $M$ is
substantially flat: this feature is in very good agreement both with
qMC computations (see \cite{Bulut_02} and references therein) and
with ARPES experiments \cite{Yoshida_01}, which detect a similar
behaviour for overdoped systems. It is worth noting that such flatness
at $X$($Y$) is present for larger dopings too ($n=0.7,$ $0.78$
and $0.85$), as shown in figure~3 of reference \cite{Avella_14a}, and
that the extension of the plateau increases upon decreasing the doping.
It is now clear that the dispersion warping along $X\to Y$, displayed
in figure~\ref{fig10}, is also responsible for two maxima in $N\left(\omega\right)$:
one related to the van-Hove singularities at $X$ and $Y$ and one
to the dispersion maximum close to $S$. The entity of the dip between
these two maxima just depends on the number of available well-defined
(red) states in $\mathbf{k}$ present between these two values of
$\omega$. This will determine the formation of a more or less pronounced
pseudogap in $N\left(\omega\right)$, as we will discuss after the
analysis of the region close to $M$.

Definitely, the absence of spectral weight close to $M$ and, in particular
and surprisingly, at $\omega=0$ (i.e., on the Fermi surface,
in contradiction with the Fermi-liquid scenario), is the most relevant
result of this study: it will determine almost all relevant and anomalous
features of the single-particle properties. Last, but not least, quite remarkable is also the presence of kinks in the dispersion in both
the nodal ($\Gamma\to M$) and the antinodal ($X\to\Gamma$) directions,
as highlighted by the dark green guidelines, in qualitative agreement
with some ARPES experiments \cite{Damascelli_03}. Remarkably, similar
results for the single-particle excitation spectrum were obtained
in the self-consistent projection operator method \cite{Kakehashi_04,Kakehashi_05},
the operator projection method \cite{Onoda_01,Onoda_01a,Onoda_03}
and a Mori-like approach by Plakida and coworkers \cite{Plakida_06,Plakida_10}.

\subsubsection{Spectral function and Fermi surface\label{sec:Spectral-Function-and}}

\begin{figure}[!b]
	\noindent \centering{}%
	\begin{tabular}{cc}
		\includegraphics[width=0.38\textwidth]{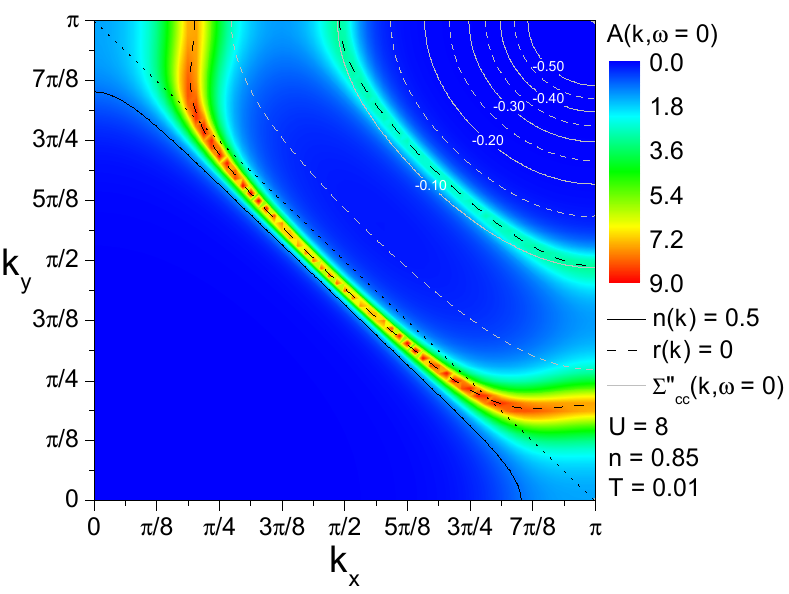} & \includegraphics[width=0.38\textwidth]{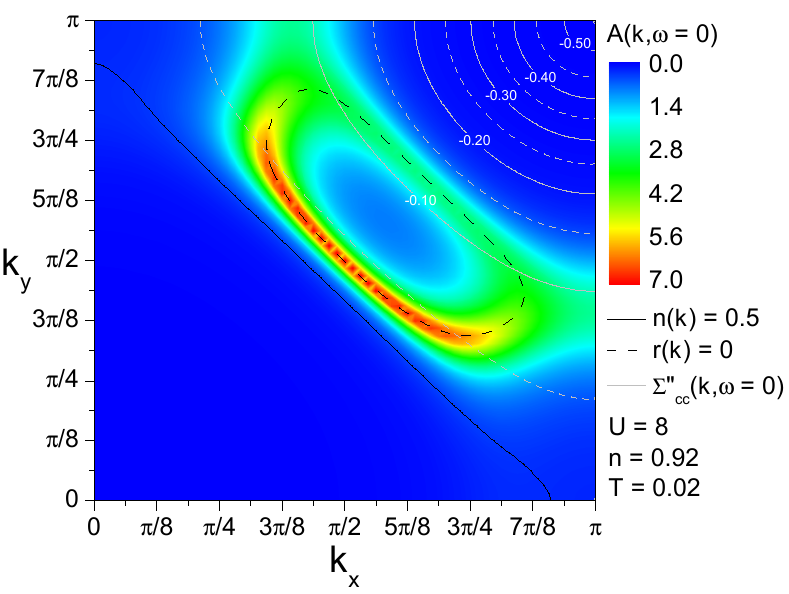}\tabularnewline
	\end{tabular}
\vspace{-2mm}
\caption{(Colour online) Spectral function at the chemical potential $A(\mathbf{k},\omega=0)$
		as a function of momentum~$\mathbf{k}$ for $U=8$, $T=0.01$ and
		(left) $n=0.85$ and (right) $n=0.92$ ($T=0.02$). The solid line
		marks the locus $n(\mathbf{k})=0.5$, the dashed line marks the locus
		$r(\mathbf{k})=0$, the gray lines are labeled with the values of
		$\Sigma''_{cc}(\mathbf{k},\omega=0)$, and the dotted line is a guide
		to the eye and marks the reduced (antiferromagnetic) Brillouin zone.
		From \cite{Avella_14a}.\label{fig20}}
\end{figure}

Considering $A(\mathbf{k},\omega=0)$, we attain the closest concept
to Fermi surface for a strongly correlated system. In figure~\ref{fig20},
we show $A(\mathbf{k},\omega=0)$ as a function of $\mathbf{k}$ in
a quarter of the Brillouin zone for $U=8$ and (left) $n=0.85$ and
$T=0.01$ and (right) $n=0.92$ and $T=0.02$. According to the interpretation
of the ARPES measurements in the sudden approximation \cite{Yoshida_01,Damascelli_03},
the Fermi surface can be defined as the locus in $\mathbf{k}$ space
of the relative maxima of $A(\mathbf{k},\omega=0)$. Such a definition
leads to a possible explanation of ARPES measurements, but also allows one
to go beyond them, with their finite resolution and sensitivity, aiming
at a reconciliation with other types of measurements, in particular
quantum oscillations ones, which seemingly report results in disagreement,
also up to dichotomy in some cases, with ARPES. 

The evolution of the Fermi surface topology on increasing $n$ is
the consequence of two Lifshitz transitions \cite{Lifshitz_60},
at $n\cong0.82$ and $n\cong0.9$. The first Lifshitz transition ($n\cong0.82$,
not shown) is due to the crossing of the chemical potential by the
van Hove singularity: the exact doping can be easily identified by
looking at the chemical potential behaviour as a function of $n$ (not
shown). The second transition ($n\cong0.9$, not shown) is characterized
by the formation of a hole pocket in the proximity of~$S$. Changes
in the Fermi-surface topology were also found in other studies on
the cuprates \cite{Trugman_90,Barabanov_01,Plakida_06}. A very detailed
analysis for the Lifshitz transitions in the $t$-$J$ model was performed
by Korshunov and Ovchinnikov \cite{Korshunov_07}: they obtained the
model parameters from \textit{ab initio} calculations for cuprates
\cite{Korshunov_05} and found a large overdoped Fermi surface for
$n<0.76$, two concentric Fermi surfaces for $0.76<n<0.85$ and a
small hole Fermi surface around $S$ for $n>0.85$. Consequently,
in reference~\cite{Ovchinnikov_11a}, Ovchinnikov~et~al. characterized
better these two quantum phase transitions, obtaining a logarithmic
singularity for $N\left(\omega=0\right)$ at $n=0.85$ and a stepwise
feature at $n=0.76$.

Then, in figure~\ref{fig20}, we report results for $n=0.85$ and $n=0.92$
as representatives of two non-trivial topologies of the Fermi surface.
We can clearly distinguish two arcs that for $n=0.92$ somehow join.
Given the current sensitivities, only the arc with the larger intensities,
among the two, may be visible to ARPES: at $n=0.92$, the region close
to $S$ is the only one with an appreciable signal. The less-intense
arc, reported in reference~\cite{Plakida_06} too, is the relic of a shadow
band, as clearly seen in figure~\ref{fig10}, and thus never changes
its curvature, in contrast to what happens to the other arc, subject
to the crossing of the van Hove singularity ($n\cong0.82$, not shown).

Three additional ingredients can help us better understand the evolution
with doping of the Fermi surface: (i) the $n(\mathbf{k})=0.5$ locus
(solid line), i.e., the Fermi surface if the system would be non-interacting;
(ii) the $r(\mathbf{k})=0$ locus (dashed line), i.e., the Fermi surface
if the system would be a Fermi liquid or a state close to it conceptually;
(iii) the values (grey lines and labels) of $\Sigma''_{cc}(\mathbf{k},\omega=0)$.
Through the combined analysis of these three ingredients with the
positions and intensities of the relative maxima of $A(\mathbf{k},\omega=0)$,
we can understand better what these latter imply and classify the
behaviour of the system on changing filling. At high dopings (not shown),
the positions of the two arcs match exactly $r(\mathbf{k})=0$ lines:
this fact corroborates our definition of Fermi surface, making it
versatile and valid beyond the Fermi liquid picture without contradicting
this latter. Increasing the filling, we find the first topological
transition from a Fermi surface closed around $\Gamma$ (hole like
in cuprates language) to a Fermi surface closed around $M$ (electron
like in cuprates language) at $n\cong0.82$ (not shown), where the
van Hove singularity is at $\omega=0$ {[}see figure~\ref{fig20} (left-hand
panel) for $n=0.85$, a close doping{]}. Close to the antinodal points
($X$ and $Y$), a net discrepancy between the position of the relative
maxima of $A(\mathbf{k},\omega=0)$ and the line $n(\mathbf{k})=0.5$
arises on decreasing doping. This feature does not only allow the
topological transition, absent for the $n(\mathbf{k})=0.5$ locus
that reaches the anti-diagonal ($X\to Y$) at $n=1$ satisfying the
Luttinger theorem, but it also accounts for the broadening of the
relative maxima of $A(\mathbf{k},\omega=0)$ close to the anti-nodal
points. The broadening is due to the small, but finite, value of $\Sigma''_{cc}(\mathbf{k},\omega=0)$
in those regions, signaling the net enhancement of the correlation
strength, and then the impossibility to consider the system in this
regime as a conventional non- (or weakly-) interacting one within
a Fermi-liquid scenario or its ordinary extensions for ordered phases.
The emergence of such features only in well defined regions in \textbf{$\mathbf{k}$}
space is of great interest and goes beyond the problem under analysis
(cuprates).

As the most important result, at $n=0.92$, the relative maxima of
$A(\mathbf{k},\omega=0)$ deviate also from the $r(\mathbf{k})=0$
line, at least partially, leading to a completely new scenario: $r(\mathbf{k})=0$
defines a pocket, while the peaks of $A(\mathbf{k},\omega=0)$ feature
the very same pocket together with quite well-defined wings closing,
with one half of the pocket, a kind of relic of a large Fermi surface.
This is the second and most surprising topological transition for
the Fermi surface; in fact, the two arcs, clearly visible for all
other dopings, join and do not close just a pocket, as expected from
the conventional theory for an antiferromagnet: they develop a fully
independent branch. The actual Fermi surface is neither a pocket nor
a large Fermi surface. This very unexpected result can be connected
to the dichotomy between the experiments (e.g., ARPES) pointing to
a small and the ones (e.g., quantum oscillations) pointing to a large
Fermi surface.

The pocket too is definitely \emph{unconventional}. In fact, there
are two distinct halves of the pocket: one with very high intensity
pinned at $S$ (the only possibly detectable by ARPES) and one with
very low intensity (detectable only by some quantum oscillations experiments).
This is our interpretation for the Fermi arcs, seen in many ARPES
experiments \cite{Damascelli_03,Shen_05} and unaccountable for any
ordinary theory based on the Fermi liquid picture, though modified
by including an incipient spin or charge ordering. Clearly, looking
only at the Fermi arc (as necessarily in ARPES), the Fermi surface
looks ill defined, not enclosing a definite region of $\mathbf{k}$
space, but having access also to the other half of the pocket, such
problem is greatly alleviated. In our understanding, the antiferromagnetic
fluctuations are so strong to destroy the quasi-particle coherence
in that region of $\mathbf{k}$ space, as similarly reported in the
DMFT$+\Sigma$ approach \cite{Sadovskii_01,Sadovskii_05,Kuchinskii_05,Kuchinskii_06,Kuchinskii_06a}
and in a Mori-like approach by Plakida and collaborators \cite{Plakida_06,Plakida_10}.
The Fermi arc is pinned at~$S$: this pinning of the \emph{center
of mass} of the ARPES-visible Fermi arc has been obtained also in
ARPES experiments~\cite{Koitzsch_04}.

\subsubsection{Density of states and pseudogap\label{sec:Density-of-States}}

\begin{figure}[!t]
	\noindent \centering{}%
	\begin{tabular}{ccccc}
		\includegraphics[height=5cm]{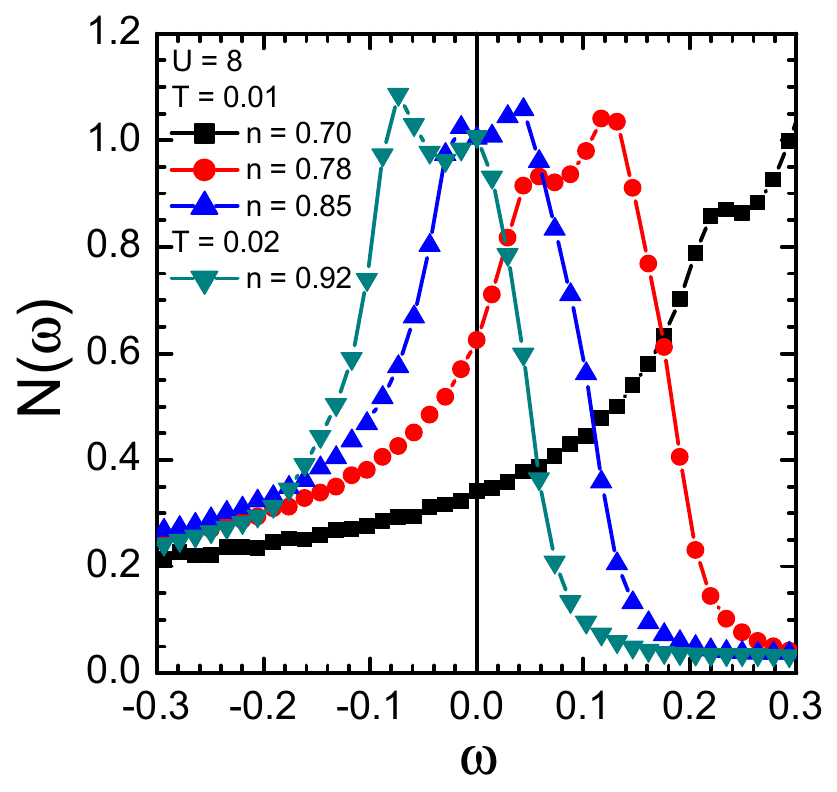} &  &  &  & \includegraphics[height=5cm]{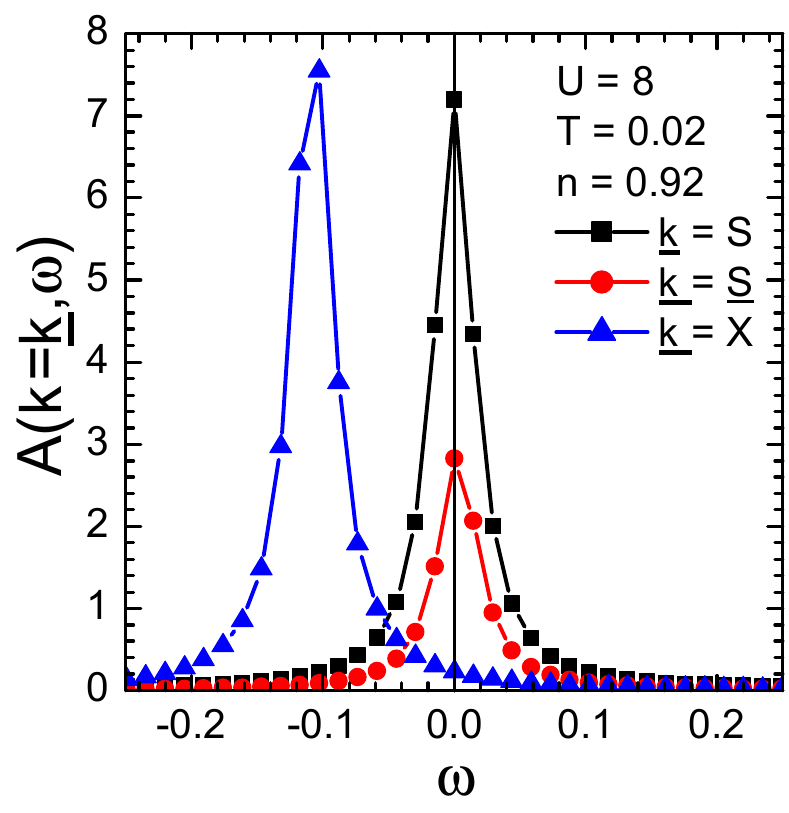}\tabularnewline
	\end{tabular}
\vspace{-2mm}
\caption{(Colour online) (left) Density of states $N(\omega)$ as a function
		of frequency $\omega$ for $U=8$, (black squares) $n=0.7$ and $T=0.01$,
		(red circles) $n=0.78$ and $T=0.01$, (blue up triangles) $n=0.85$
		and $T=0.01$, and (green down triangles) $n=0.92$ and $T=0.02$.
		(right) Spectral function in proximity of the chemical potential $A(\mathbf{\underline{k}},\omega\sim0)$
		at (black squares) $\mathbf{\underline{k}}=S=(\piup/2,\piup/2)$, (red
		circles) $\underline{S}$ (in the text), and (blue triangles) $X=(\piup,\,0)$
		for $U=8$, $n=0.92$ and $T=0.02$. From \cite{Avella_14a}.\label{fig8}}
\end{figure}

The other main feature in underdoped cuprates is a substantial depletion
in the electronic $N\left(\omega\right)$: the \textit{pseudogap}.
In figure~\ref{fig8} (left-hand panel), we show $N(\omega)$ for $U=8$
and four couples of values of filling and temperature: $n=0.7$ and
$T=0.01$, $n=0.78$ and $T=0.01$, $n=0.85$ and $T=0.01$, and $n=0.92$
and $T=0.02$, in the frequency region close to the chemical potential.
As a reference, we also show, in figure~\ref{fig8} (right-hand panel),
$A(\mathbf{\underline{k}},\omega\sim0)$ at $\mathbf{\underline{k}}=S=(\piup/2,\piup/2)$,
$\mathbf{\underline{k}}=\underline{S}$ which is where the \emph{phantom}
half of the pocket touches the diagonal $\Gamma\to M$ (i.e., where
the dispersion cuts the diagonal $\Gamma\to M$ closer to $M$), and
$\mathbf{\underline{k}}=X$ for $U=8$, $n=0.92$ and $T=0.02$.
As apparent in figure~\ref{fig8} (left-hand panel), $N\left(\omega\right)$
has two maxima separated by a dip, which plays the role of pseudogap.
Its presence is due to the dispersion warping along the $X\to Y$
direction (see figure~\ref{fig10}), which induces the two maxima {[}one
due to the van-Hove singularity at $X$ and one due to the dispersion
maximum close to $S$ --- see figure~\ref{fig8} (right-hand panel){]} and
the loss of states, in this frequency window, in the region in $\mathbf{k}$
close to $M$. In order to realize the weight loss due to the finite
value of $\Sigma''_{cc}(\mathbf{k},\omega=0)$ in the region in $\mathbf{k}$
close to $M$, we can look at the impressive difference between the
values of $A(\underline{S},\omega=0)$ and $A(S,\omega=0)$ {[}figure~\ref{fig8}
(right-hand panel){]}. Increasing the filling, there is a net spectral
weight transfer between the two maxima; in particular, from the dispersion
top close to $S$ to the antinodal point $X$, where the van Hove
singularity resides. At the lowest doping ($n=0.92$), a well formed
pseudogap can be seen for $\omega<0$ and will clearly influence the
properties of the system. For $n=0.92$, we do not find any divergence
of $\Sigma'_{cc}(\mathbf{k},\omega=0)$, in contrast to what is stated
in reference~\cite{Stanescu_05} where this feature is presented as the
definite reason for the pseudogap formation. In our study, the pseudogap
is just the result of the weight transfer from the single-particle
density of states to the two-particle one, related to the (antiferro)magnetic
excitations occurring in the system on decreasing doping at low $T$.
A similar doping behaviour of the pseudogap has been found by the DMFT$+\Sigma$
approach \cite{Sadovskii_01,Sadovskii_05,Kuchinskii_05,Kuchinskii_06,Kuchinskii_06a},
a Mori-like approach by Plakida and collaborators \cite{Plakida_06,Plakida_10}
and the cluster perturbation theory \cite{Senechal_00,Senechal_04,Tremblay_06}.

\section{Conclusions}

In this review paper, written in honor of the $80^{\textrm{th}}$ anniversary
of Professor Ihor Stasyuk, we have illustrated the composite operator
method by means of a prototypical example: the 2D Hubbard model and
its application to the description of high-$T_\text{c}$ cuprate superconductors.
The method has been reported first for a general case in order to appreciate
its philosophy and its capability to be applied to any strongly correlated
system in a controlled fashion.


\vspace{-2mm}

\ukrainianpart 

\title{Підхід методу композитних операторів до двовимірної моделі Хаббарда та 
	купратів} 
\vspace{-2mm}
\author{А. Ді Чьоло\refaddr{label1}, А. Авелла\refaddr{label1,label2,label3}} 
\addresses{ 
	\addr{label1} факультет фізики ``Е.Р. Caianiello'', університет Салерно, 
	I-84084 Фішіано (Салерно), Італія 
	\addr{label2} CNR-SPIN, університет Салерно, I-84084 Фішіано (Салерно), Італія 
	\addr{label3} відділення CNISM в Салерно, університет Салерно, I-84084 Фішіано (Салерно), Італія 
}

\vspace{-5mm}

\makeukrtitle 

\begin{abstract} 
	\tolerance=3000%
	У цій оглядовій статті ми показуємо можливий шлях до отримання вірогідних 
	розв'язків для двовимірної моделі Хаббарда та пояснення деякої незвичної 
	поведінки недолегованих високотемпературних купратних надпровідників в рамках 
	методу композитних операторів. Сам метод описано вичерпно в його 
	фундаментальній філософії, різних інгредієнтах та надійній техніці для 
	з'ясування причин його успішного застосування до різноманітних сильно 
	скорельованих систем, суперечливих феноменологій та загадкових матеріалів. 
	\keywords багаточастинкові методи, сильно скорельовані системи, модель 
	Хаббарда, купрати, псевдощілина, електронна структура 
	
\end{abstract}

\lastpage
\end{document}